# Role of chiral quantum Hall edge states in nuclear spin polarization

Kaifeng Yang[1], Katsumi Nagase[2], Yoshiro Hirayama[2], Tetsuya Mishima[3], Michael Santos[3] & Hongwu Liu[1]

[1]State Key Lab of Superhard Materials and Institute of Atomic and Molecular Physics, Jilin University, Changchun 130012, P. R. China

[2]Department of Physics, Tohoku University, Sendai, Miyagi 980-8578, Japan

[3]Homer L. Dodge Department of Physics and Astronomy, University of Oklahoma, 440 West Brooks, Norman, OK 73019-2061, USA

Correspondence and requests for materials should be addressed to H.W.L. (email: hwliu@jlu.edu.cn).

Resistively detected NMR (RDNMR) based on dynamic nuclear polarization (DNP) in a quantum Hall ferromagnet (QHF) is a highly-sensitive method for the discovery of fascinating quantum Hall phases; however, the mechanism of this DNP and in particular the role of quantum Hall edge states in it are unclear. Here we demonstrate the important but previously unrecognized effect of chiral edge modes on the nuclear spin polarization. A side-by-side comparison of the RDNMR signals from Hall bar and Corbino disk configurations allows us to distinguish the contributions of bulk and edge states to DNP in QHF. The unidirectional current flow along chiral edge states makes the polarization robust to thermal fluctuations at high temperatures and makes it possible to observe a reciprocity principle of the RDNMR response. These findings help us better understand complex NMR responses in QHF, which has important implications for the development of RDNMR techniques.

Resistively detected NMR (RDNMR)[1] developed in a quantum Hall ferromagnet (QHF)[2-4] of GaAs two-dimensional electron gases (2DEGs) at filling factor $\nu = 2/3$ (corresponding to a composite-fermion filling factor $\nu^* = 2$)[5,6] has been widely used to discover exotic 2D electronic states[7-10] and to coherently control the nuclear spins in 2DEGs[11]. The RDNMR technique depends on the current-induced dynamic nuclear polarization (DNP) that is expected to occur by transferring spin polarization from electrons to nuclei via the electron-nuclear hyperfine interaction at a domain wall (DW) separating two energetically-degenerate domains[12]. This polarization process is in contrast to DNP by electron spin resonance[1] or optical[13-15] pumping of intra- or inter-band transitions to generate nonequilibrium electron spin polarizations which then polarize the nuclei via the hyperfine interaction during subsequent relaxation to equilibrium. However, the detailed mechanism is still poorly understood. In particular, the above-mentioned studies of the $\nu = 2/3$ QHF are performed on the Hall bar where contributions from both bulk and edge states to DNP[16-19] coexist and also the edge physics at $\nu = 2/3$ remains unclear[20], which may complicate its interpretation.

The $\nu = 2/3$ QHF is classified as an easy-axis ferromagnet according to its magnetic anisotropy energy[4]. Such ferromagnetic ground states have also been formed in integer QH regimes of various 2DEGs[21-25]. It is known that QH edge states at integer $\nu$ correspond to bulk Landau levels (LLs) below the Fermi energy[26], which are chiral in the sense that they propagate in only one direction on a given edge of a Hall bar – the right-moving state on the top edge and the left-moving one on the bottom edge (or vice versa, depending on the orientation of magnetic field). The chiral edge states are immune to backscattering and localization provided there is no interedge scattering[27], which accounts for a non-dissipative (quantized) transport in the QH effect[28].

Here we focus on the edge state in the dissipative transport of QHF where its chiral character has received little attention and show how chiral modes establish DNP. We present comparative RDNMR measurements of the simplest easy-axis QHF at $\nu = 2$ of InSb 2DEGs[29] patterned into Hall bar and Corbino disk configurations. The absence of edge states in the Corbino disk allows us to investigate DNP in bulk[16], which provides a basis for discussion of DNP via edge states of the Hall bar. This side-by-side comparison experiment reveals a reciprocity principle of the NMR response in the $\nu = 2$ QHF at temperatures where the bulk contribution to DNP vanishes but the intraedge-scattering-induced DNP still operates, highlighting the important role of chiral edge states on DNP. Our results clearly show that the chiral edge state has direct effects on the nuclear spin polarization besides its known effects on the electron transport in QH systems.

**Results**

**RDNMR measurements of the Corbino disk.** The magnetoresistance and RDNMR measurements of InSb 2DEGs were performed in a dilution refrigerator (see Methods). A comparative RDNMR study was carried out on the $\nu = 2$ QHF of the two configurations that was a highly sensitive region for the detection of DNP (see Methods), focusing on the dependence of RDNMR signals on the type [alternating current (AC) or direct current (DC)] of electric current, the direction of current flow, the orientation of magnetic field, and the effect of temperature. We first present the results obtained from the Corbino disk. Figure 1 shows that the DC RDNMR signal of $^{115}$In has a dispersive line shape (DLS) with quadrupole splittings at low temperatures and disappears at 2 K. Note that the signals at or below 1 K are nearly independent of temperature due to the current-induced heating[30]. Figure 1b depicts the domain structures of QHF, where the spin-polarized (spin polarization $P = 1$) and spin-unpolarized ($P = 0$) domains are separated by DW (order of the magnetic length $l_B$ in width[31]). Charge transport across the DW accompanying electron spin flip between two energetically-degenerate domains is responsible for an emerging conductivity spike that characterizes the QHF (Supplementary Note 1). Although compared to the hyperfine interaction the spin-orbit coupling (large in InSb) is much more efficient to flip electron spins, the role played by nuclear spins is still evident from the observed RDNMR signal that is generally accepted to be caused by DNP at DW boundaries via electron-nuclear flip-flop[12]. The interplay between the hyperfine and spin-orbit coupling in the RDNMR sensitivity deserves future study. The electric field gradient at nuclear positions induced by strain[19] in the InSb QW accounts for the quadrupole coupling to ten nuclear energy levels of $^{115}$In with nuclear spin $I_N = 9/2$[32]. Single-photon transitions among these levels are expected to result in nine quadrupole resonances with equal frequency intervals. Because the conductivity change $\Delta\sigma_{xx}$ (see Methods) is determined by a population ($N$) of each nuclear energy level that depends on the spin configuration of electrons coupled to nuclei[33], we propose that the nuclear population profiles near the $P = 1$ and $P = 0$ domains are different (Fig. 1c). The population difference $\Delta N$ (Fig. 1c,d) will increase (decrease) the Zeeman splitting $E_Z$ of electron spins in the $P = 1$ ($P = 0$) domain via the Overhauser effect, which is equivalent to increasing (decreasing) a parallel field but keeping the perpendicular component $B_{perp}$ constant in a tilted-magnetic-field measurement because the Overhauser effect has no influence on electron orbital motion determined by $B_{perp}$. This results in a leftward (rightward) shift of the conductive spike after the DNP[29] and thus a dip (peak) in the RDNMR signal. The DLS in data suggests that the polarized nuclear spins near the $P = 1$ and $P = 0$ domains separated by a domain size (several hundreds of nm[34]) give the same weight to the RDNMR response as all possible electron trajectories across the DW between the two sides (AC and BD) are involved (called bulk mode). This interpretation is further supported by the fact that the peak-to-dip pattern is reversed as the signal is taken from the other side of the spike (data not shown). From the above discussion it follows that the DC RDNMR signal is independent of the direction of current flow (Fig.1a) and the orientation of magnetic field (Supplementary Fig. 1). In addition, the DC signal is found to be cool-down independent (Supplementary Fig. 2).

In contrast to the DC measurement where the polarized nuclei with opposite spins are well separated, the current flow with opposite directions in the AC measurement produces the two species at both sides of one DW that are gradually distributed over a narrow width[12]. In this case, the nuclear spin polarization is negligible on average that is responsible for the absence of RDNMR signals (Supplementary Fig. 3). The role of AC current is also played by high temperatures in the DC measurement, where thermal fluctuations make the direction of electron transfer between neighboring domains[35] opposite to that driven by the DC current and thus suppress the RDNMR signal (Fig. 1a).

**RDNMR measurements of the Hall bar.** The results obtained from the Corbino disk provide a good reference for the RDNMR study of the Hall bar where both bulk and edge states coexist. The temperature-dependent DC RDNMR spectrum of the Hall bar is shown in Fig. 2a, which is found to be significantly different from that of the Corbino disk. In particular, the signals are present at high temperatures up to 6 K with changes in line shape. For ease of comparison, we summarize the results of a detailed analysis of the signals of both configurations in Fig. 2b,c. It is clear that the two configurations have a DLS with relatively symmetric peak and dip heights for both directions of current flow at or below 1 K despite the large difference in signal amplitude. The signal amplitude of the Corbino disk goes directly to zero as the temperature is raised to 2 K, while that of the Hall bar decreases rapidly with increasing temperature first and then shows a gradual decrease from 3 to 6 K. Furthermore, the signal asymmetry ratio in this temperature range is large and almost constant with the sign depending on the direction of current flow. These findings lead us to conclude that DNP is dominated by the bulk mode at low temperatures and the presence of edge states in the Hall bar is responsible for the observed differences (see below). Figure 2d depicts the domain structures of QHF in the Hall bar, where the edge states become part of an array of domains. It is seen that the two edges of the sample are connected by channels along domain boundaries, which is supported by optically detected magnetic resonance imaging of the $\nu = 2/3$ QHF of GaAs 2DEGs[20]. The edge transport will affect the bulk mode of the Hall bar as follows: the chiral nature of edge states determines that electrons feeding into the sample from one side (e.g., point A) either go back to the same side (point C) by travelling along the DW without spin flip or reach the other side (point B) by passing across the DW with spin flip[36]. Although the former process does not contribute to DNP directly, it tends to reduce the number of electrons passing across the DW that not only results in the spike with high resistance (or $\sigma_{xx}$, Supplementary Fig. 4) but also improves the RDNMR sensitivity as indicated by a large signal amplitude in Fig. 2b. Furthermore, the edge transport is also responsible for the dependence of signal amplitude of the Hall bar on the direction of current flow (Fig. 2b), the orientation of magnetic field (Supplementary Fig. 1), and different cool-downs (Supplementary Fig. 2), as discussed in Supplementary Note 2.

**Reciprocity principle of the RDNMR signal.** A distinct change in the temperature dependence of signal amplitude and asymmetry ratio of the Hall bar suggests that the mechanism of DNP may change from the bulk mode to the edge mode (i.e., DNP due to electron-nuclear flip-flop via edge states with opposite spins as indicated by thick dashed arrows in Fig. 2d, also called intraedge-scattering-induced DNP). This is further supported by the observation of a reciprocity principle of the RDNMR response in Fig. 3. For the edge-state picture of the QH effect, the chiral nature (i.e., the one-way electron motion) makes a difference in current between the top and bottom edge states when a driving current is induced, as denoted by line thickness in Fig. 3a,b,g,h. This difference makes the DNP mainly occur along either of the two spatially separated paths of lane 1 and lane 2, resulting in different signal line shapes in Fig. 3c-f. The signal line shape is found to be unchanged provided the lane mainly responsible for DNP is kept the same by simultaneously reversing both current flow and magnetic field (Fig. 3a,b,g,h), while the direction of electron motion along the same lane only changes the signal amplitude. Note that because the reversal of current difference between the two lanes is compatible with that of a Hall voltage, we can also say that the reciprocity principle in $\Delta\sigma_{xx}$ similar to the Hall voltage occurs in the edge dominated region. It follows from the signal line shape and the above discussion that the polarized nuclear spins near the $P = 1$ ($P = 0$) domain give more weight to the RDNMR response as electrons move along the lane 1 (lane 2). The chiral nature also ensures that electron motion along the edge state is unidirectional and thus robust to thermal fluctuations, accounting for the presence of RDNMR signals at high temperatures. It is worth noting that the signal amplitude determined by the thermally robust edge mode is small and less sensitive to temperature (Fig. 2b), indicating that the edge mode is overwhelmed by the bulk mode at low temperatures. Therefore, we conclude that DNP is dominated by the bulk mode at low temperatures but by the edge mode at high temperatures when the bulk mode is completely suppressed. This edge mode is also present in the AC RDNMR measurement of the Hall bar where DNP in bulk is suppressed (Supplementary Note 2).

**Discussion**

The observed reciprocity principle of the NMR response also helps us better understand DLS of the $\nu = 2$ QHF. We infer from Fig. 3 that electron trajectories in the bulk mode can be regarded as convergence of the two lanes in the edge mode zig-zagging throughout the 2D plane with the same transmission probability and current flow direction but without the chiral nature, resulting in DLS. As discussed above, the DNP-induced decrease (increase) of the electronic Zeeman energy near the $P = 0$ ($P = 1$) domain results in the peak (dip) feature and the difference between peak and dip heights is determined by the weight of these two domains given to the NMR response. Furthermore, as compared with the $\Delta N$-$f$ dependence in Fig. 1d, the emergence of the 5$^{th}$ resonance line in data is suggestive of the Knight shift ($K_s$) of the RDNMR response near the $P = 1$ domain relative to that near the $P = 0$ domain. The frequency spacing between peak and dip of this resonance gives $K_s \sim 35$ kHz. Note that, because the quadrupole splitting $\Delta f \sim 85$ kHz is much larger than $K_s$ in our case, the frequency spacing between peak and dip of the DLS as a whole is determined by the quadrupole splittings ($\sim 5\Delta f$) rather than by $K_s$. With this understanding, we now proceed to investigate the possible role of domain structures in the $\nu = 1$ DLS of the GaAs 2DEG whose origin is a long standing mystery[37-43]. The most recent research[44,45] confirms that the peak and dip signals are attributed to the coupling of nuclei to spin-unpolarized and spin-polarized 2DEGs, respectively, and the frequency spacing between them is determined by $K_s$ (this determination is consistent with ours made at the 5$^{th}$ resonance line). To explain these findings, it is further proposed that the 2DEG near $\nu = 1$ could spontaneously break symmetry to form domains with polarized and unpolarized regions. However, we have to note that in contrast to the $\nu = 2$ QHF where the current-induced DNP induces opposite changes in the electronic Zeeman energy of different domains that is responsible for DLS, the assigned thermal nuclear polarization near $\nu = 1$[37] can only decrease the electronic Zeeman energy and result in a dip. Thus, the NMR-induced heating of the 2DEG together with the Zeeman effect has been considered[40] but ruled out by recent experiments[44-46]. One possibility to account for the $\nu = 1$ DLS using the domain scenario is that the current applied to $\nu = 1$ with domain structures in the literature might be large enough to induce DNP that exceeds thermal nuclear polarization. Further studies following our work are required to examine this possibility.

Finally, we note that our understanding of the role of chiral edge states in DNP and complex NMR responses in the simplest easy-axis QHF at $\nu = 2$ may shed light on the study of DNP in the $\nu = 2/3$ (or $\nu^* = 2$) QHF. In addition, the RDNMR spectra at $\nu = 2/3$ have recently been used to determine the topology of various QH phases (stripe, bubble, Wigner and Skyrme crystals)[9,10,47,48], where the Knight shift of all nuclei that have considerable overlap with the electron wave function is summed to calculate the signal line shape. We emphasize that attention should be paid to the spatial distribution of DNP varying with sample configurations and experimental conditions as discussed in our work, which may affect the profile of subband wavefunctions at filling factors used for the signal readout and thus the NMR line shape.

**Methods**

**Sample preparation and characterization**. The 2DEG in a 20-nm-wide InSb quantum well grown on GaAs (001) substrates[49] (Supplementary Fig. 5c) was patterned into Corbino disk and Hall bar configurations (Supplementary Fig. 5a,b) simultaneously on one chip that was subjected to the same measurement procedures for our comparison purpose. Indium was used for Ohmic contacts in both samples. The Corbino disk was defined by two circular Ohmic contacts (source S and drain D) with radii of $r_1 = 95$ μm and $r_2 = 195$ μm, respectively, and the Hall bar had a length of $L = 100$ μm and a width of $W = 30$ μm. The following 2DEG parameters were measured on the Hall bar at $T = 100$ mK in a dilution refrigerator equipped with in-situ rotator stage using a standard AC lock-in technique at 13.7 Hz. Determination of the tilt angle $\theta$ between the sample normal and $B$ (Supplementary Fig. 4, inset) was made by measuring the slope of low-field Hall resistance. The electron mobility ($\mu$) and density ($n_s$) were obtained from the Hall measurement at $\theta = 0°$ and found to be cool-down dependent [e.g., $\mu = 20$ (20.6) m$^2$/Vs and $n_s = 2.66$ (2.7) × $10^{15}$ m$^{-2}$ for the first (second) cool-down]. The effective mass $m^* \sim 0.016$ in units of the free-electron mass $m_e$ was determined by analyzing the temperature-dependent amplitude of low-field Shubnikov-de Haas oscillations[50]. The coincidence technique was used to measure the product of $m^*g^*$ at the LL intersection and thus the effective $g$-factor $g^*$ that was shown to be linear with spin polarization of each LL intersection[50]. The QHF spike studied here was formed at the $\nu = 2$ LL intersection with $\theta = 64.1°$ (Supplementary Fig. 4), where $g^* \sim 55$ was obtained.

**RDNMR measurement**. A low-noise preamplifier (Stanford Research Systems, Model SR560) and a standard AC lock-in technique at 13.7 Hz were used for the DC and AC RDNMR measurements, respectively, at temperatures from 100 mK to 6 K. The RDNMR measurements were performed on the $\nu = 2$ QHF formed at the energy gap $\varepsilon = 0$ (Supplementary Fig. 4, inset). The details of the RDNMR measurement are as follows: a large current is applied to polarize the nuclei around the $\nu = 2$ spike, as indicated by an exponential increase in $\sigma_{xx}$ on a time scale of hundreds of seconds (Supplementary Fig. 6a). After $\sigma_{xx}$ becomes saturated ($\sigma_{xx}^{sat}$), a continuous-wave radio-frequency (RF) field (~ μT) at a power of 0 dBm generated by a single turn coil surrounding the sample is applied to irradiate the 2DEG. The change in $\sigma_{xx}$ with respect to $\sigma_{xx}^{sat}$ during frequency ($f$) sweep through the resonance condition of $f_{NMR} = \gamma B$ ($\gamma$, the gyromagnetic ratio of $^{115}$In) defines $\Delta\sigma_{xx}$, which describes the depolarization of nuclei. A slow sweep rate (12 kHz/min) is used in order that $\sigma_{xx}$ at each frequency point approaches the equilibrium value. The $f$ dependence of $\Delta\sigma_{xx}/\sigma_{xx}^{sat}$ represents the RDNMR spectrum. Note that an increase in $n_s$ for the second cool-down results in a shift of the $\nu = 2$ spike towards higher magnetic fields and thus a difference in the field strength (12 T and 12.3 T for the first and second cool-downs, respectively) at which the RDNMR measurement is performed.

**Detection sensitivity**. As the gap $\varepsilon$ is made to approach zero by adjusting $\theta$ (Supplementary Fig. 4, inset), $\varepsilon$ and thus the $\varepsilon = 0$ position are strongly influenced by the hyperfine contribution to the electronic Zeeman splitting, $\Delta_{HF} = \Sigma_j A^{(j)} <I^{(j)}>$ (where $A$ and $<I>$ are the hyperfine interaction constant and nuclear spin polarization of different nuclei $j$, respectively[1]). This results in a shift of the QHF spike and allows detection of the RDNMR signal, which is similar to the RDNMR measurement of the $\nu = 2/3$ QHF in the GaAs 2DEG[7]. We calculate that the two systems have a comparable $\Delta_{HF}$ (446 μeV in InSb and 140 μeV in GaAs) if all nuclei are fully polarized (i.e., $<I^{(j)}>$ is equal to the nuclear spin $I_N$ of each nuclear isotope). A relatively large $\Delta_{HF}$ in InSb is mainly due to the large $I_N$. It is estimated that the degree of nuclear polarization ($P_N$) in the $\nu = 2$ QHF of the InSb 2DEG is about 10% (see below), which is comparable to that in the $\nu = 2/3$ QHF of the GaAs 2DEG[51]. This degree of polarization results in $\Delta\sigma_{xx}/\sigma_{xx}^{sat}$ on the order of a few percent. From the above discussion it follows that QHF is a highly-sensitive region for the RDNMR measurement. Here we note that the effective nuclear field $B_N = \Delta_{HF}/(g^*\mu_B)$ in InSb is extremely small due to large $g^*$: an absolute value $|B_N|$ in InSb with $g^* \sim -55$ is only about 0.14T if $\langle I^{(j)} \rangle = I_N$ while that in GaAs with $g^* \sim -0.44$ is $\sim 5.3$T. In our study, $|B_N| \sim 0.014$T is calculated by $B_N = B_{perp}/\cos(\theta + \Delta\theta) - B_{perp}/\cos\theta$, where $\Delta\theta$ is an equivalent change in angle due to DNP that is deduced from $\Delta\theta = \left.\frac{\Delta\sigma_{xx} d\theta}{d\sigma_{xx}}\right|_{B_{perp}}$ with $\frac{d\sigma_{xx}}{d\theta}$ obtained from the angle dependence of the spike position[29]. Therefore, we have $P_N = 0.014\text{T}/0.14\text{T} = 10\%$.

**Data availability.** The data that support the findings of this study are available from the corresponding author upon request.

## References

1. Dobers, M., von Klitzing, K., Schneider, J. & Ploog, K. Electrical detection of nuclear magnetic resonance in GaAs-Al$_x$Ga$_{1-x}$As heterostructures. *Phys. Rev. Lett.* **61**, 1650-1653 (1988).

2. Pisazza, V., Pellegrini, V., Beltram, F., Wegscheider, W., Junwirth, T. & MacDonald, A. H. First-order phase transitions in a quantum Hall ferromagnet. *Nature* **402**, 638-641 (1999).

3. Poortere, E. P., Tutuc, E., Papadakis, S. J. & Shayegan, M. Resistance spikes at transitions between quantum Hall ferromagnets. *Science* **290**, 1546-1549 (2000).

4. Jungwirth, T. & MacDonald, A. H. Pseudospin anisotropy classification of quantum Hall ferromagnets. *Phys. Rev. B* **63**, 035305 (2000).

5. Kronmüller, S., Dietsche, W., Weis, J. & von Klitzing, K. New resistance maxima in the fractional quantum Hall effect regime. *Phys. Rev. Lett.* **81**, 2526-2529 (1998).

6. Kronmüller, S., Dietsche, W., von Klitzing, K., Denninger, G., Wegscheider, W. & Bichler M. New type

of electron nuclear-spin interaction from resistively detected NMR in the fractional quantum Hall effect regime. *Phys. Rev. Lett.* **82**, 4070-4073 (1999).

7. Smet, J. H., Deutschmann, R. A., Ertl, F., Wegscheider, W., Abstreiter, G. & von Klitzing, K. Gate-voltage control of spin interactions between electrons and nuclei in a semiconductor. *Nature* **415**, 281-286 (2002).

8. Kumada, N., Muraki, K. & Hirayama, Y. Low-frequency spin dynamics in a canted antiferromagnet. *Science* **313**, 329-332 (2006).

9. Tiemann, L., Gamez, G., Kumada, N. & Muraki, K. Unraveling the spin polarization of the ν = 5/2 fractional quantum Hall state. *Science* **335**, 828-831 (2012).

10. Tiemann, L., Rhone, T. D., Shibata, N. & Muraki, K. NMR profiling of quantum electron solids in high magnetic fields. *Nat. Phys.* **10**, 648-652 (2014).

11. Yusa, G., Muraki, K., Takashina, K., Hashimoto, K. & Hirayama, Y. Controlled multiple quantum coherences of nuclear spins in a nanometer-scale device. *Nature* **434**, 1001-1005 (2005).

12. Kumada, N., Kamada, T., Miyashita, S., Hirayama, Y. & Fujisawa, T. Electric field induced nuclear spin resonance mediated by oscillating electron spin domains in GaAs-based semiconductors. *Phys. Rev. Lett.* **101**, 137602 (2008).

13. Barrett, S. E., Tycko, R., Pfeiffer, L. N. & West, K. W. Directly detected nuclear magnetic resonance of optically pumped GaAs quantum wells. *Phys. Rev. Lett.* **72**, 1368-1371 (1994).

14. Tycko, R., Barrett, S. E., Dabbagh, G., Pfeiffer, L. N. & West, K. W. Electronic states in gallium arsenide quantum wells probed by optically pumped NMR. *Science* **268**, 1460-1463 (1995).

15. Akiba, K., Kanasugi, S., Yuge, T., Nagase, K. & Hirayama, Y. Optically induced nuclear spin polarization in the quantum Hall regime: the effect of electron spin polarization through exciton and trion excitations. *Phys. Rev. Lett.* **115**, 026804 (2015).

16. Kawamura, M, Takahashi, H., Masubuchi, S., Hashimoto, Y., Katsumoto, S., Hamaya, K. & Machida, T. Dynamic nuclear polarization in a quantum Hall Corbino disk. *J. Phys. Soc. Jpn* **77**, 023710 (2008).

17. Wald, K. R., Kouwenhoven, L. P., McEuen, P. L., van der Vaart, N. C. & Foxon, C. T. Local dynamic nuclear polarization using quantum point contacts. *Phys. Rev. Lett.* **73**, 1011-1014 (1994).

18. Machida, T., Yamazaki, T. & Komiyama, S. Local control of dynamic nuclear polarization in quantum Hall devices. *Appl. Phys. Lett.* **80**, 4178-4180 (2002).

19. Bowers, C. R., Caldwell, J. D., Gusev, G., Kovalev, A. E., Olshanetsky, E., Reno, J. L., Simmons, J. A. & Vitkalov, S. A. Dynamic nuclear polarization and nuclear magnetic resonance in the vicinity of edge states of a 2DES in GaAs quantum wells. *Solid State Nucl. Magn. Reson.* **29**, 52-65 (2006).

20. Moore, J. N., Hayakawa, J., Mano, T., Noda, T. & Yusa, G. Non-equilibrium fractional quantum Hall states visualized by optically detected MRI. Preprint at http://arxiv.org/abs/1606.06416 (2016).

21. Jaroszyński, J., Andrearczyk, T., Karczewski, G., Wróbel, J., Wojtowicz, T., Papis, E., Kamińska, E., Piotrowska, A., Popović, D. & Dietl, T. Ising quantum Hall ferromagnet in magnetically doped quantum wells. *Phys. Rev. Lett.* **89**, 266802 (2002).

22. Chokomakoua, J. C., Goel, N., Chung, S. J., Santos, M. B., Hicks, J. L., Johnson, M. B. & Murphy, S. Q. Ising quantum Hall ferromagnetism in InSb-based two-dimensional electronic systems. *Phys. Rev. B* **69**, 235315 (2004).

23. Vakili, K., Shkolnikov, Y. P., Tutuc, E., Bishop, N. C., De Poortere, E. P. & Shayegen, M. Spin-dependent resistivity at transitions between integer quantum Hall states. *Phys. Rev. Lett.* **94**, 176402 (2005).

24. Young, A. F., Dean, C. R., Wang, L., Ren, H., Cadden-Zimansky, P., Watanabe, K., Taniguchi, T., Hone,

J., Shepard, K. L. & Kim, P. Spin and dip quantum Hall ferromagnetism in graphene. *Nat. Phys.* **8**, 550-556 (2012).

25. Yakunin, M. V., Suslov, A. V., Podgornykh, S. M., Dvoretsky, S. A. & Mikhailov, N. N. Effect of spin polarization in the HgTe quantum well. *Phys. Rev. B* **85**, 245321 (2012).

26. Halperin, B. I. Quantized Hall conductance, current-carrying edge states, and the existence of extended states in a two-dimensional disordered potential. *Phys. Rev. B* **25**, 2185-2190 (1982).

27. Büttiker, M. Absence of backscattering in the quantum Hall effect in multipleprobe conductors. *Phys. Rev. B* **38**, 9375-9389 (1988).

28. von Klitzing, K., Dorda, G. & Pepper, M. New method for high-accuracy determination of the fine-structure constant based on quantized Hall resistance. *Phys. Rev. Lett.* **45**, 494-497 (1980).

29. Liu, H. W., Yang, K. F., Mishima, T. D., Santos, M. B. & Hirayama, Y. Dynamic nuclear polarization and nuclear magnetic resonance in the simplest pseudospin quantum Hall ferromagnet. *Phys. Rev. B* **82**, 241304 (R) (2010).

30. Yang, K. F., Liu, H. W., Nagase, K., Mishima, T. D., Santos, M. B. & Hirayama, Y. Resistively detected nuclear magnetic resonance via a single InSb two-dimensional electron gas at high temperatures. *Appl. Phys. Lett.* **98**, 142109 (2011).

31. Jungwirth, T. & MacDonald, A. H. Resistance spikes and domain wall loops in Ising quantum Hall ferromagnets. *Phys. Rev. Lett.* **87**, 216801 (2001).

32. Dzhioev, R. I. & Korenev, V. L. Stabilization of the electron-nuclear spin orientation in quantum dots by the nuclear quadrupole interaction. *Phys. Rev. Lett.* **99**, 037401 (2007).

33. Ota, T., Yusa, G., Kumada, N., Miyashita, S. & Hirayama, Y. Nuclear spin population and its control toward initialization using an all-electrical submicron scale nuclear magnetic resonance device. *Appl.*

*Phys. Lett.* **90**, 102118 (2007).

34. Verdene, B. Martin, J., Gamez, G., Smet, J., von Klitzing, K., Mahalu, D., Schuh, D., Abstreiter, G. & Yacoby, A. Microscopic manifestation of the spin phase transition at filling factor 2/3. *Nat. Phys.* **3**, 392-396 (2007).

35. Kang, P. G., Jeong, H. & Yeom, H. W. Hopping domain wall induced by paired adatoms on an atomic wire: Si (111)-(5×2)-Au. *Phys. Rev. Lett.* **100**, 146103 (2008).

36. Mitra, A. & Girvin, S. M. Electron/nuclear spin domain walls in quantum Hall systems. *Phys. Rev. B* **67**, 245311 (2003).

37. Desrat, W., Maude, D. K., Potemski, M., Portal, J. C., Wasilewski, Z. R. & Hill, G. Resistively detected nuclear magnetic resonance in the quantum Hall regime: possible evidence for a Skyrme crystal. *Phys. Rev. Lett.* **88**, 256807 (2002).

38. Stern, O., Freytag, N., Fay, A., Dietsche, W., Smet, J. H., von Klitzing, K., Schuh, D. & Wegscheider, W. NMR study of the electron spin polarization in the fractional quantum Hall effect of a single quantum well: spectroscopic evidence for domain formation. *Phys. Rev. B* **70**, 075318 (2004).

39. Gervais, G., Stormer, H. L., Tsui, D. C., Engel, L. W., Kuhns, P. L., Moulton, W. G., Reyes, A. P., Pfeiffer, L. N., Baldwin, K. W. & West, K. W. NMR in the solid phase of two-dimensional electrons at high magnetic fields. *Phys. Rev. B* **72**, 041301(R) (2005).

40. Tracy, L. A., Eisenstein, J. P., Pfeiffer, L. N. & West, K. W. Resistively detected NMR in a two-dimensional electron system near $v = 1$: clues to the origin of the dispersive lineshape. *Phys. Rev. B* **73**, 121306(R) (2006).

41. Kodera, K., Takado, H., Endo, A., Katsumoto, S. & Iye, Y. Dispersive lineshape of the resistively-detected NMR in the vicinity of Landau level filling $v = 1$. *Phys. Stat. Sol. C* **3**, 4380-4383

(2006).

42. Dean, C. R., Piot, B. A., Gervais, G., Pfeiffer, L. N. & West, K. W. Current-induced nuclear-spin activation in a two-dimensional electron gas. *Phys. Rev. B* **80**, 153301 (2009).

43. Bowers, C. R., Gusev, G. M., Jaroszynski, J., Reno, J. L. & Simmons, J. A. Resistively detected NMR of the $v = 1$ quantum Hall state: a tilted magnetic field study. *Phys. Rev. B* **81**, 073301 (2010).

44. Desrat, W., Piot, B. A., Krämer, S., Maude, D. K., Wasilewski, Z. R., Henini, M. & Airey, R. Dispersive line shape in the vicinity of the $v = 1$ quantum Hall state: coexistence of Knight-shifted and unshifted resistively detected NMR responses. *Phys. Rev. B* **88**, 241306(R) (2013).

45. Desrat, W., Piot, B. A., Maude, D. K., Wasilewski, Z. R., Henini, M. & Airey, R. W line shape in the resistively detected nuclear magnetic resonance. *J. Phys.: Condens. Matter* **27**, 275801 (2015).

46. Yang, K. F., Liu, H. W., Mishima, T. D., Santos, M. B., Nagase, K. & Hirayama, Y. Resistively detected NMR with dispersive lineshape in single InSb quantum wells. *J. Phys.: Conf. Ser.* **334**, 012029 (2011).

47. Rhone, T. D., Tiemann, L. & Muraki, K. NMR probing of spin and charge order near odd-integer filling in the second Landau level. *Phys. Rev. B* **92**, 041301 (R) (2015).

48. Côté, R. & Simoneau, A. M. Resistively detected NMR spectra of the crystal states of the two-dimensional electron gas in a quantizing magnetic field. *Phys. Rev. B* **93**, 075305 (2016).

49. Uddin, M. M., Liu, H. W., Yang, K. F., Nagase, K., Mishima, T. D., Santos, M. B. & Hirayama, Y. Characterization of InSb quantum wells with atomic layer deposited gate dielectrics. *Appl. Phys. Lett.* **101,** 233503 (2012).

50. Yang, K. F., Liu, H. W., Mishima, T. D., Santos, M. B., Nagase, K. & Hirayama, Y. Nonlinear magnetic field dependence of spin polarization in high-density two-dimensional electron systems. *New J. Phys.* **13**, 083010 (2011).

51. Miyamoto S., Hatano T., Watanabe S. & Hirayama Y. NMR tracing of hyperfine-mediated nuclear spin diffusion in fractional quantum Hall domain phases. Preprint at http://arxiv.org/abs/1605.06926 (2016).

**Acknowledgements**

This work was supported in part by the grants Program for New Century Excellent Talents of University in China (H.W.L.), Scientific Research Foundation of Jilin University for the Overseas Scholar (No.419080500395, K.F.Y.), JST-ERATO, KAKENHI (No. 26287059 and No. 15H05867, Y.H.), WPI-AIMR in Tohoku University (Y.H.), and GP-Spin in Tohoku University (K.N. and Y.H.).

**Author contributions**

K.F.Y. carried out measurements and collected the data. K.N. fabricated InSb Hall bars and Corbino disks. T.D.M. and M.B.S. grew InSb heterostructures. Y.H. contributed to helpful discussions. H.W.L. conceived and designed the experiments. K.F.Y. and H.W.L. analyzed the data and wrote the paper and all co-authors commented on it.

**Additional information**

**Supplementary Information** accompanies this paper at http://www.nature.com/naturecommunications

**Competing financial interests:** The authors declare no competing financial interests.

**Figure 1 | Temperature dependence of direct current (DC) RDNMR spectra of $^{115}$In in a Corbino disk.**
(**a**) $\Delta\sigma_{xx}/\sigma_{xx}^{\mathrm{sat}}$ versus $f$ as a function of temperature at $B$ = 12.3 T with current $I$ = 0.6 μA (black curve) and $I$ = -0.6 μA (pink cruve). The dash-dotted line represents the zero level. Quadrupole resonances are indicated by vertical solid lines with numbers 1-9. (**b**) Schematic domain structures of quantum Hall ferromagnet (QHF). The gray and green areas denote the spin-unpolarized [spin polarization $P = 0$, spin-up (black solid arrow) and spin-down (red solid arrow) electrons in the two Zeeman levels of the $n = 0$ Landau level (LL), see inset of Supplementary Fig. 4] and spin-polarized ($P = 1$, spin-up electrons in both $n = 0$ and $n = 1$ LLs) domains, respectively, and a domain wall (DW) occurs in between. For clarity, spin-up electrons in the $n = 0$ LL are not shown in the graph. The electron-spin flip (say from spin-down to spin-up, red dashed arrows) flops one nuclear spin from spin-up (black hollow arrow) to spin-down (red hollow arrow) at DW boundaries. Note that the arrow length is not scaled with the magnetic moment of each particle. (**c**) A possible population distribution (energy $E$ versus population $N$) of $^{115}$In with ten nuclear spin states $|m>$ near the $P = 0$ and $P = 1$ domains and the total population distribution by assigning the same weight from these two domains to the RDNMR response. The presence of electric quadrupole coupling accounts for a difference in the splitting between these levels (where $f_0$ and $\Delta f$ are the Zeeman and quadrupole frequencies, respectively, and $h$ is Planck's constant). (**d**) The corresponding population difference between adjacent levels (denoted by numbers 1-9), $\Delta N = N_{|m>} - N_{|m-1>}$, as a function of $f$, where the frequency interval is equally spaced by $\Delta f$ and the largest $\Delta N$ is taken as unity. The total $\Delta N$-$f$ dependence is proposed to be responsible for quadrupole resonances in data that are equally spaced by $\Delta f \sim 85$ kHz.

**Figure 2 | Temperature dependence of DC RDNMR spectra of $^{115}$In in a Hall bar.** (**a**) $\Delta\sigma_{xx}/\sigma_{xx}^{\text{sat}}$ versus $f$ as a function of temperature at $B = 12.3$ T with $I = 1$ μA (black curve) and $I = -1$ μA (pink curve). The dash-dotted line represents the zero level. (**b,c**) Temperature dependence of the amplitude (peak-to-dip height) of $\Delta\sigma_{xx}/\sigma_{xx}^{\text{sat}}$ and the asymmetry ratio of the height difference between peak and dip to the amplitude of $\Delta\sigma_{xx}/\sigma_{xx}^{\text{sat}}$ obtained from the data of **a** and Fig. 1. The dotted lines in **b** are guides for the eye. (**d**) Schematic

domain structures of QHF. The top and bottom lines denote the edge state corresponding to the pseudospin-up $[(n, \sigma) = (0,\uparrow)]$ LL and the solid line surrounding domains denotes the one corresponding to either the pseudospin-down $[(n, \sigma) = (0,\downarrow)]$ or pseudospin-up $[(n, \sigma) = (1,\uparrow)]$ LL. Line thickness represents the relative intensity of edge current. Electron-nuclear flip-flop via the bulk (edge) state is indicated by thin (thick) dashed arrows.

**Figure 3 | Reciprocity principle of the RDNMR response in a Hall bar**. DC RDNMR spectra of $^{115}$In measured at $T = 3$ K, $B = \pm 12.3$ T and $I = \pm 1$ μA in **c-f** together with the corresponding schematic domain structures of QHF in **a,b,g,h** show that the RDNMR line shape depends on which edge current path mainly contributes for DNP: one between points A and B (called lane 1) and the other between points C and D (lane 2). This reciprocal RDNMR response is obtained by requiring both current flow and magnetic field to be reversed. The dash-dotted line represents the zero level.

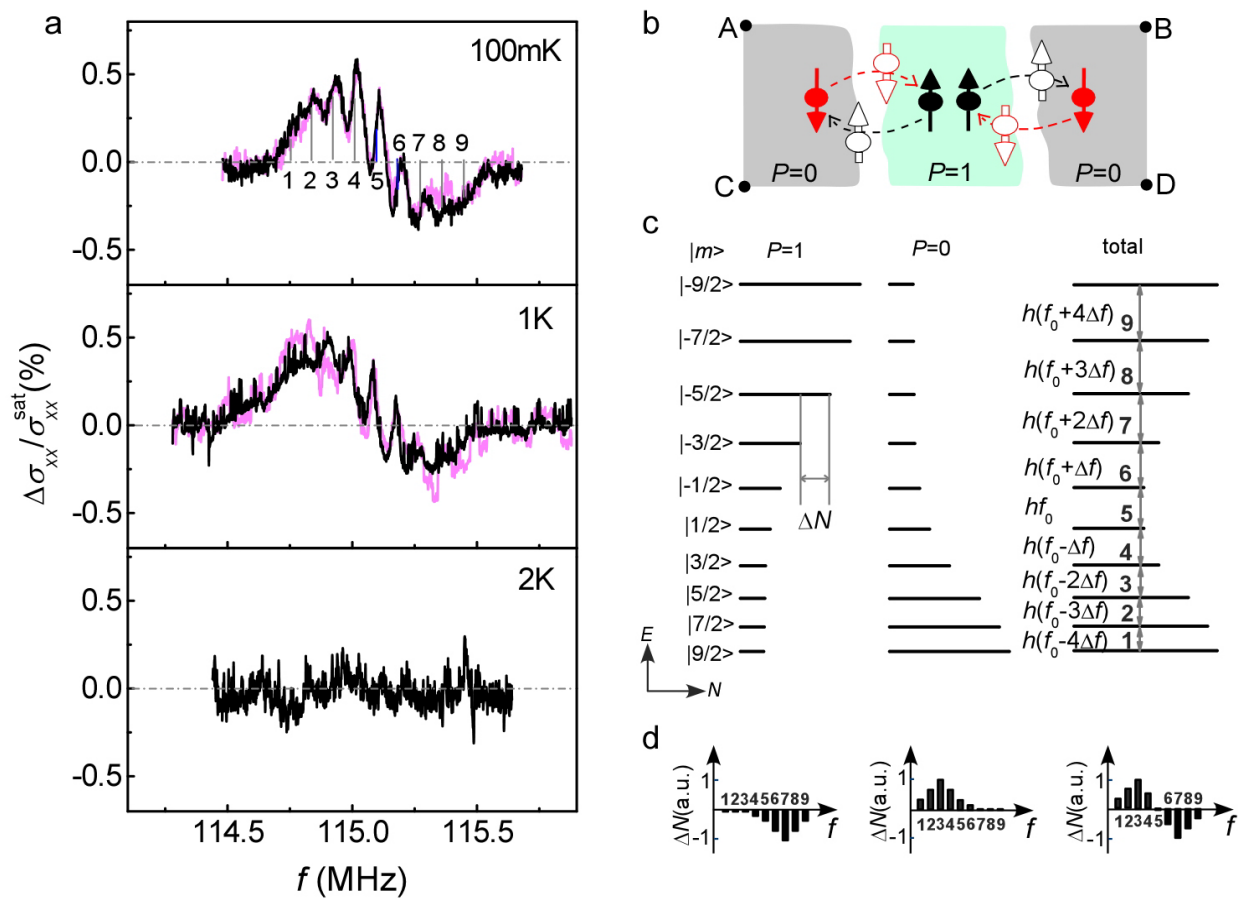

**Figure 1**

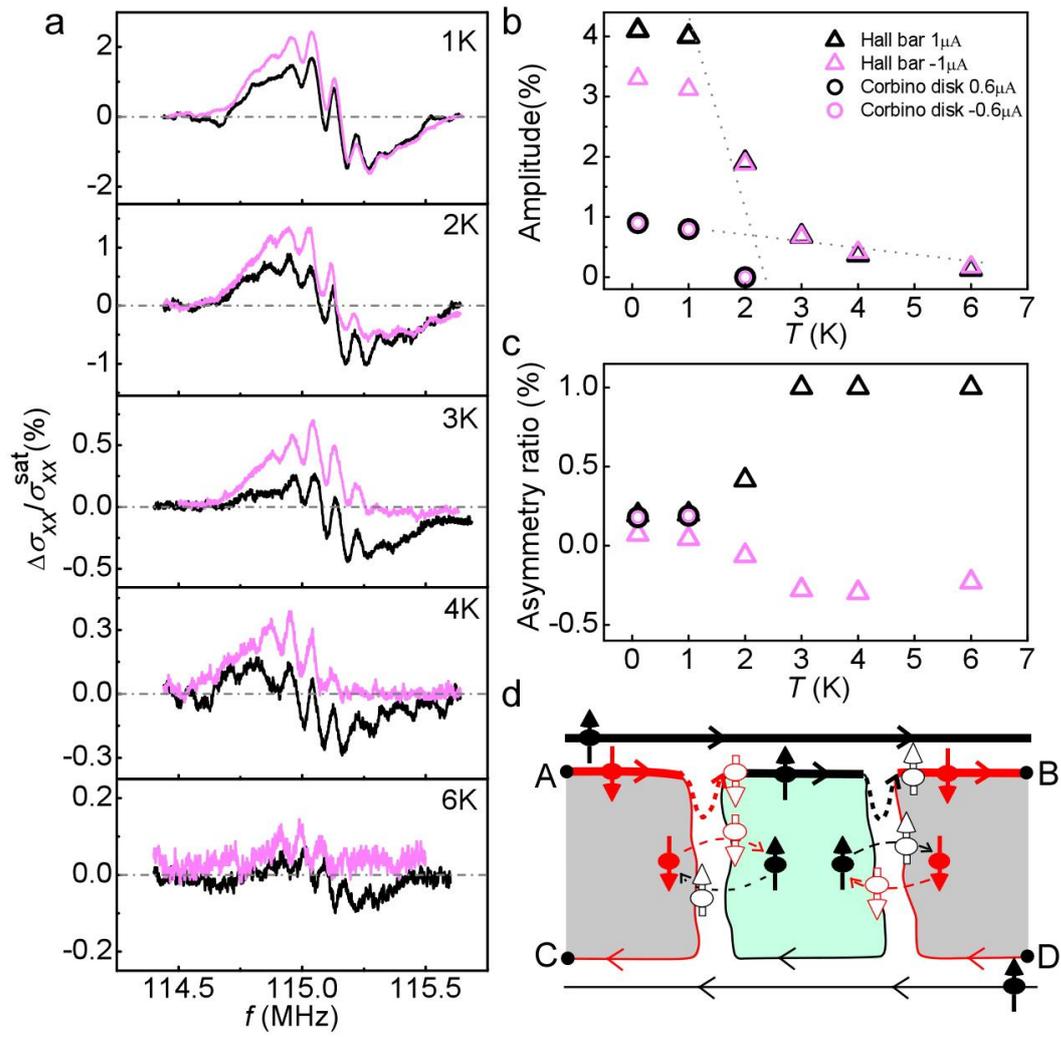

**Figure 2**

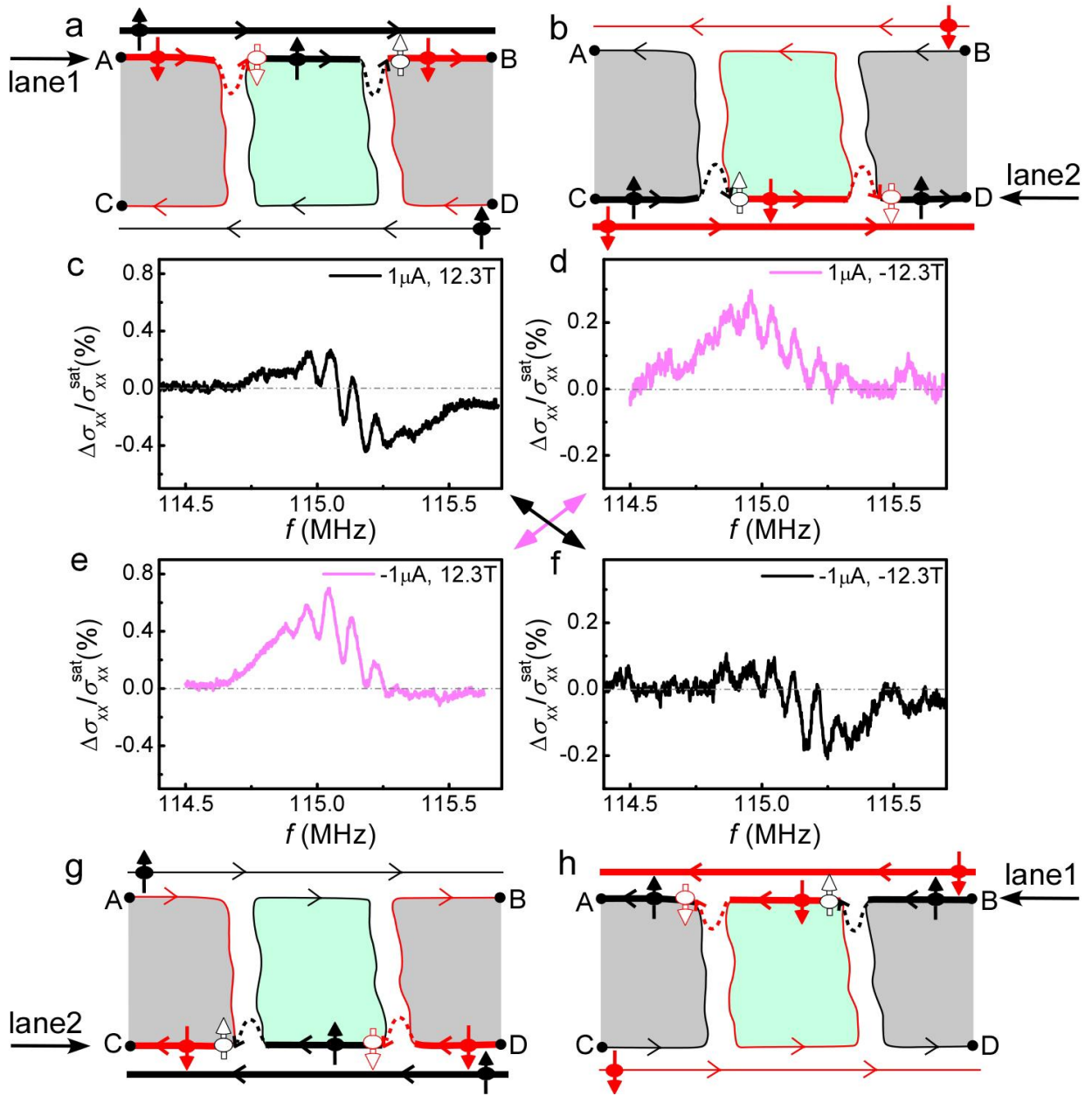

**Figure 3**

# Supplementary Information

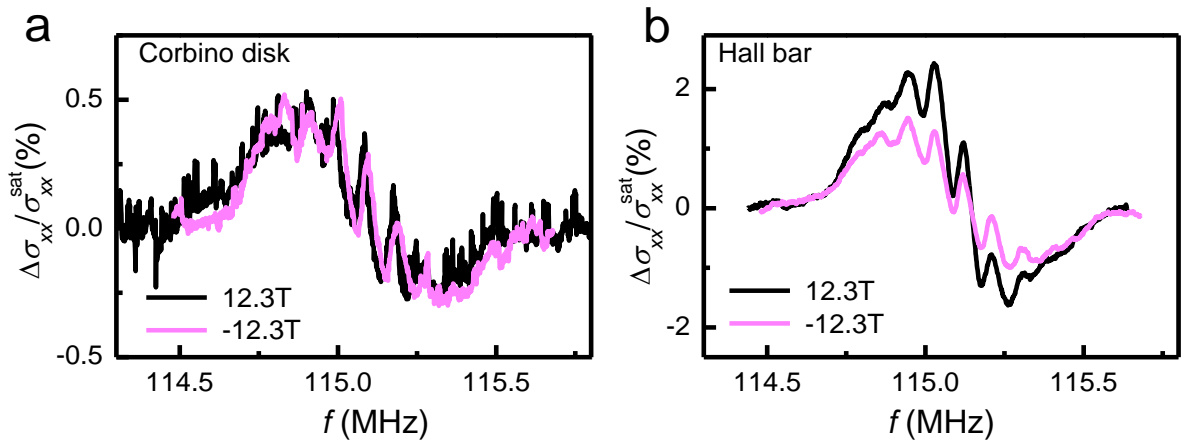

**Supplementary Figure 1: Direct current (DC) RDNMR signals at different *B* orientations.** $\Delta\sigma_{xx}/\sigma_{xx}^{\text{sat}}$ versus *f* for both Corbino disk (**a**) and Hall bar (**b**) at $B = \pm 12.3$ T and $T = 100$ mK.

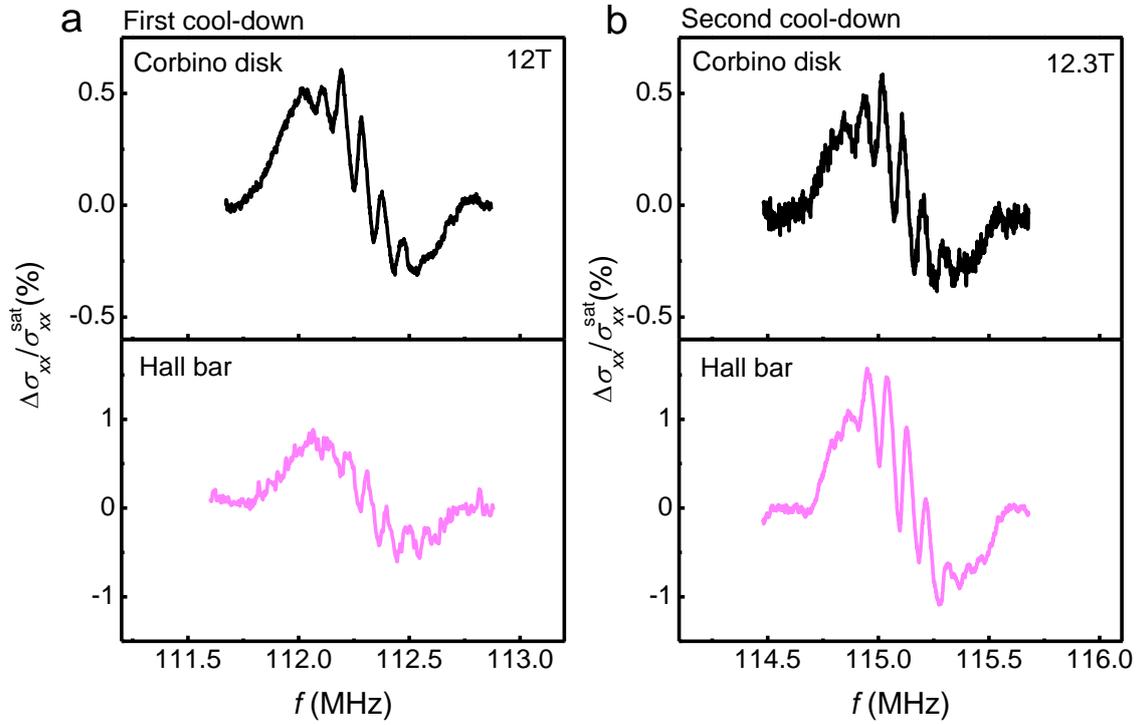

**Supplementary Figure 2: DC RDNMR signals for different cool-downs.** $\Delta\sigma_{xx}/\sigma_{xx}^{sat}$ versus $f$ for both Corbino disk and Hall bar measured after the first (**a**) and second (**b**) cool-downs to $T = 100$ mK. Note that the amplitude scale is different for the two samples.

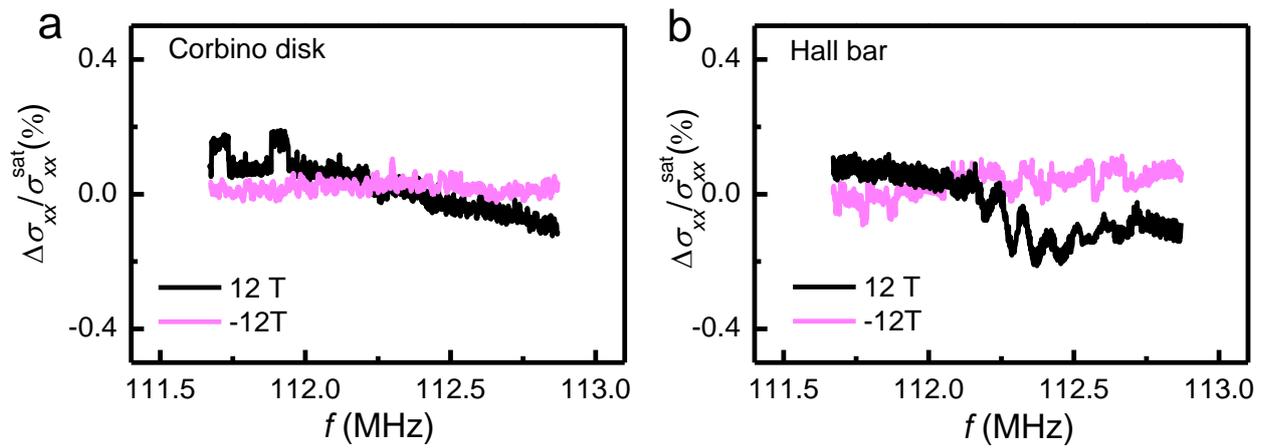

**Supplementary Figure 3: Alternating current (AC) RDNMR signals at different *B* orientations.**

$\Delta\sigma_{xx}/\sigma_{xx}^{sat}$ versus *f* for both Corbino disk (**a**) and Hall bar (**b**) at $B = \pm 12$ T and $T = 100$ mK.

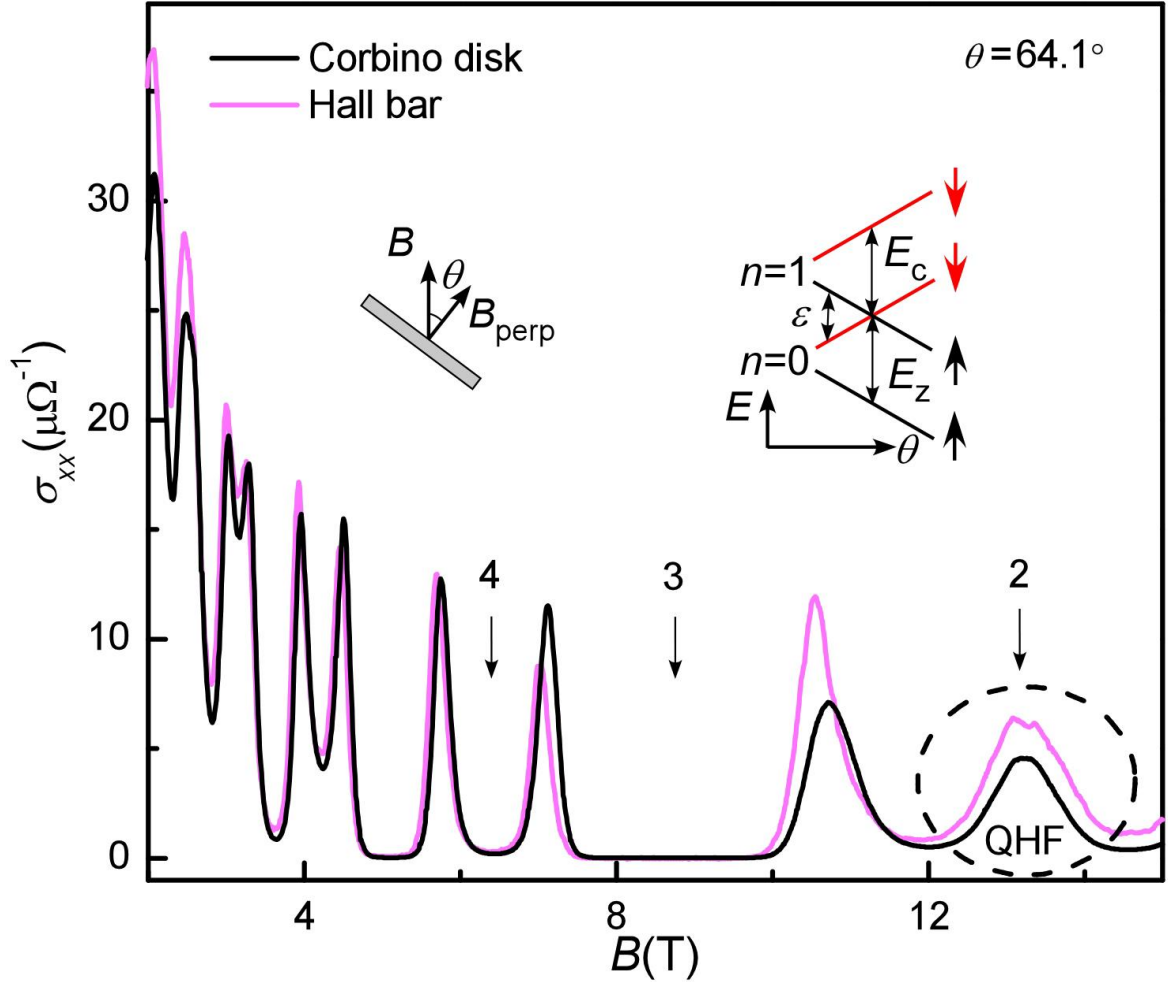

**Supplementary Figure 4: Quantum Hall ferromagnet (QHF).** $\sigma_{xx}$ versus $B$ for both Corbino disk and Hall bar with a tilt angle of $\theta = 64.1°$ ($\theta$ is the angle between $B_{perp}$ along the sample normal and $B$ shown in the left inset) measured after the first cool-down to $T = 100$ mK. The conductivity spike at $\nu = 2$ marked by circles is a signal of QHF that is formed at the LL intersection with the energy gap $\varepsilon = 0$ when the Zeeman splitting of $E_z = g^* \mu_B B = g^* \mu_B B_{perp} / \cos\theta$ (where $\mu_B$ is the Bohr magneton) and the cyclotron splitting of $E_c = \hbar e B_{perp} / m^*$ (where $\hbar$ is the reduced Planck's constant and $e$ is the electron charge) are made equal by adjusting $\theta$ (right inset).

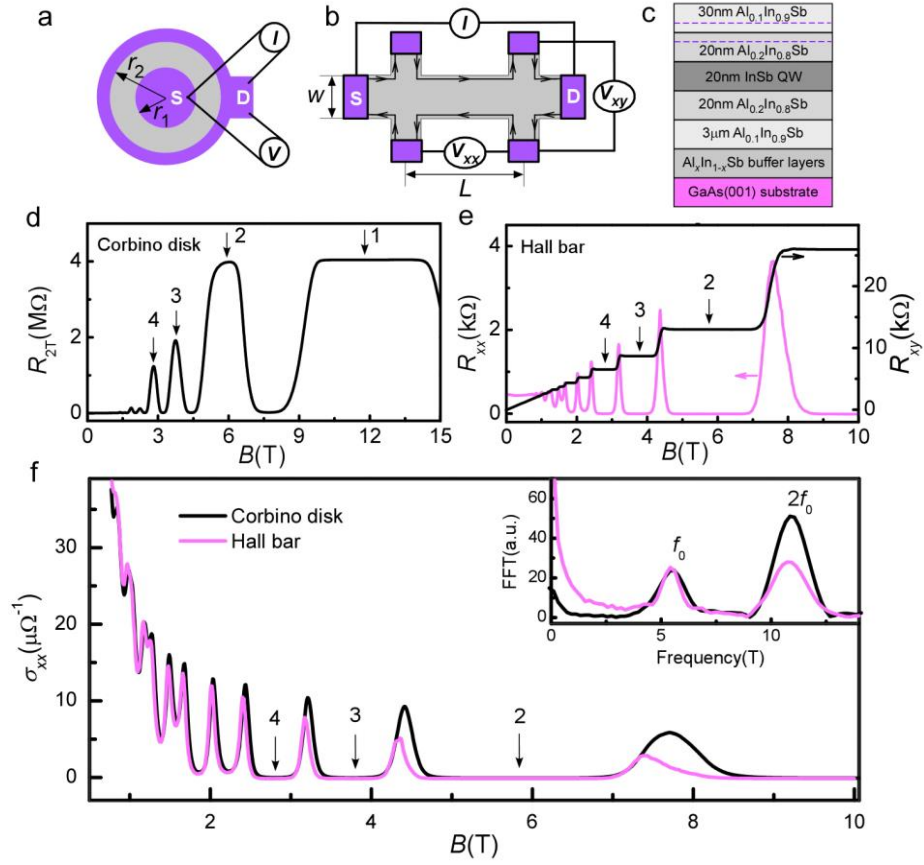

**Supplementary Figure 5: Quantum Hall effect.** Corbino disk (**a**) and Hall bar (**b**) configurations fabricated on the two-dimensional electron gas (2DEG) in an InSb quantum well (**c**). The Corbino disk is defined by two circular Ohmic contacts (source S and drain D) with radii of $r_1 = 95$ μm and $r_2 = 195$ μm, respectively. The S-D current ($I$) and output voltage ($V$) give a two-terminal resistance $R_{2T}$. The Hall bar has a length of $L = 100$ μm and a width of $W = 30$ μm. The S-D current $I$ and output voltage $V_{xx}$ ($V_{xy}$) give longitudinal resistances $R_{xx}$ (Hall resistances $R_{xy}$). The trajectory of electrons carried by edge states along the sample boundary whose direction is determined by the orientation of $B$ is shown in **b**. **d**, $R_{2T}$ versus $B$. The number indicates the Landau-level (LL) filling factor $\nu$. **e**, $R_{xx}$ and $R_{xy}$ versus $B$. **f**, $\sigma_{xx}$ versus $B$ for both Corbino disk and Hall bar calculated by $\sigma_{xx} = \ln(r_2/r_1)/(2\pi R_{2T})$ and $\sigma_{xx} = \rho_{xx}/(\rho_{xx}^2 + \rho_{xy}^2)$ ($\rho_{xx} = R_{xx}W/L; \rho_{xy} = R_{xy}$), respectively. Inset shows the fast Fourier transform (FFT) spectrum taken from the low-field SdH oscillations of the two configurations, from which $n_s = 2ef_0/h = 2.66 \times 10^{15}\,\text{m}^{-2}$ is obtained. All measurements were performed after the first cool-down to $T = 100$ mK.

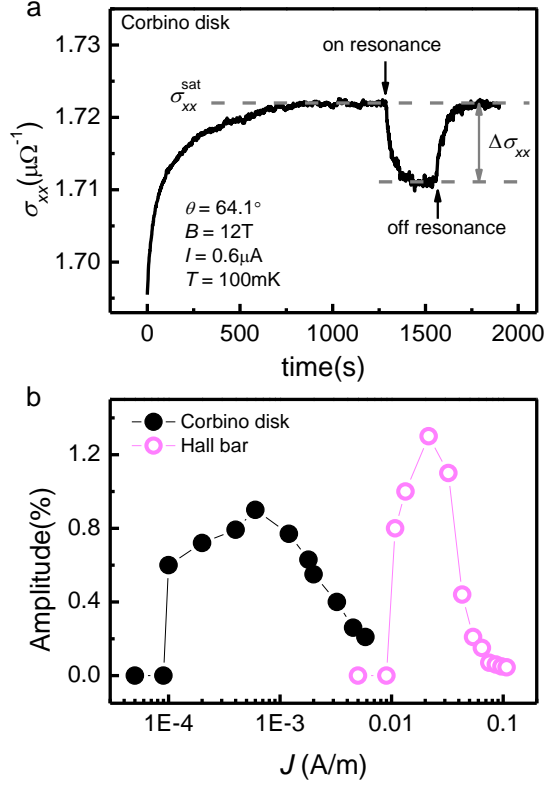

**Supplementary Figure 6: Current dependence of RDNMR signal amplitude. a**, Time dependence of $\Delta\sigma_{xx}$ in the Corbino disk. The "on resonance" ("off resonance") condition corresponds to the radio frequency ($f$) matching (mismatching) the gyromagnetic ratio of $^{115}$In. **b**, Amplitude (peak to dip) of $\Delta\sigma_{xx}/\sigma_{xx}^{sat}$ versus the DC density ($J$) for both Corbino disk ($J = I/2\pi r_1$) and Hall bar ($J = I/W$) at $B = 12$ T and $T = 100$ mK. Because $J$ has a radial dependence in the Corbino disk, the calculation at inner circular contacts gives an upper bound. Note that a decrease in amplitude with increasing $J$ for both samples is due to the nuclear depolarization caused by the current-induced heating.

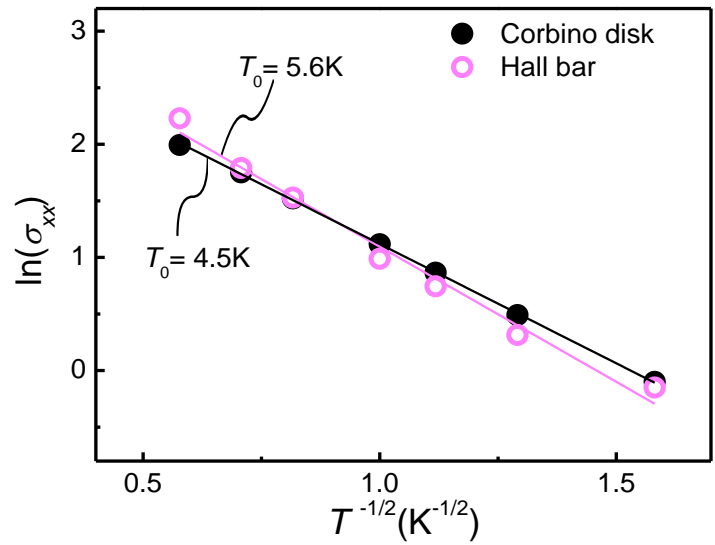

**Supplementary Figure 7: Variable-range hopping transport.** Temperature dependence of $\sigma_{xx}$ at $B = 12$ T around which the RDNMR measurement was performed (Supplementary Fig.4). The lines are the fits to the variable-range hopping formula $\sigma_{xx} \propto \exp\left(-\sqrt{T_0/T}\right)$ ($T_0$, the characteristic temperature).

**Supplementary Note1: Magnetotransport in the quantum Hall and quantum Hall ferromagnet regimes**

We examined the quantum Hall effect and magnetotransport properties of both Corbino disk and Hall bar using the alternating current (AC) measurement in a magnetic field ($B$) perpendicular to the sample substrate. It is shown in Supplementary Fig. 5d,e that the Corbino disk has high resistance $R_{2T} \sim$ MΩ near an integer filling factor $\nu$ where the Hall bar has zero longitudinal resistance $R_{xx}$ with the Hall plateau. Electrons in the Corbino disk are trapped in localized bulk states at integer $\nu$ that makes it electrically insulating, while those in the Hall bar are carried by either edge channel or Hall current path in bulk that accounts for a non-dissipative transport. Although the two samples have different magnetoresistance properties, they have a similarity in diagonal conductivity $\sigma_{xx}$ (Supplementary Fig. 5f) that is most directly related to the Landau-level (LL) nature and characterizes the bulk dissipation near the Fermi energy[1]. The $\sigma_{xx}$-$B$ dependence at low fields is found to be consistent with each other, suggesting the same electron mobility and density in both samples. Note that an asymmetric line shape of the $\sigma_{xx}(B)$ peak for the Hall bar at high fields is caused by the difference in coupling efficiency between the edge and bulk states as LLs are raised above the Fermi level with increasing $B$[2]. This does not occur for the Corbino disk without edge states.

We constructed and characterized the quantum Hall ferromagnet (QHF) of both Corbino disk and Hall bar using the AC measurement in tilted magnetic fields. The QHF is formed at $\nu = 2$ where the pseudospin-down [$(n, \sigma) = (0,\downarrow)$] and pseudospin-up [$(n, \sigma) = (1,\uparrow)$] LLs ($n$ and $\sigma$ are the orbital and spin indices, respectively) are brought into degeneracy (i.e., the single-particle energy gap $\varepsilon$ is zero) by modifying the tilt angle $\theta$ to balance the Zeeman and cyclotron splitting (inset, Supplementary Fig. 4)[3]. Strong electron exchange interactions at the LL intersection favor the formation of the simplest pseudospin QHF, in which disorder or finite temperature produces a domain wall (DW) separating the two domains with different pseudospin polarizations[4,5]. Charge transport across the DW is assigned to account for an emerging peak (so-called spike, marked by circle in Supplementary Fig. 4) within the persistent conductivity minima that is the signature of QHF. The line shape and position of the spike are found to be the same for both samples, while a large spike amplitude for the Hall bar is caused by edge transport across the DW (see main text).

The temperature dependence of conductivity minima in the QHF region is shown in Supplementary Fig. 7, which is well fitted with variable-range hopping theory[6]. We expect that hopping transport between charged topological defects (Skyrmion-like) trapped in DW[4] dominates the dissipative process as the network of DWs forms a percolation cluster through the entire 2D plane[7]. It is shown that the derived characteristic temperature $T_0$ ($\propto 1/\xi$, $\xi$ is the localization length) in the two samples is similar, suggesting that the domain/DW morphology is independent of sample geometries. This helps us to demonstrate the role of edge states in dynamic nuclear polarization (DNP) by a comparative resistively detected NMR (RDNMR) study.

**Supplementary Note 2: Direct current (DC) and AC RDNMR measurements**

The temporal evolution of $\sigma_{xx}$ after a large current is applied to polarize the nuclei around the QHF spike is shown in Supplementary Fig. 6a. The RDNMR measurement is performed when $\sigma_{xx}$ becomes saturated ($\sigma_{xx}^{sat}$) (see Methods) and the RF frequency ($f$) dependence of $\Delta\sigma_{xx}/\sigma_{xx}^{sat}$ represents the RDNMR spectrum. It is shown in Supplementary Fig. 6b that the critical current density needed for the DC RDNMR measurement of the Corbino disk is much smaller than that of the Hall bar. This suggests that the current flowing along the lowest edge state of the Hall bar (corresponding to the pseudospin-up [($n$, $\sigma$) = (0,↑)] LL, inset of Supplementary Fig. 4) is much larger than the current flowing across the DW that contributes to DNP. In the comparative RDNMR study, we set the operating current to approach the maximum amplitude of $\Delta\sigma_{xx}/\sigma_{xx}^{sat}$ in each sample ($I$ = 0.6 μA for the Corbino disk and $I$ = 1 μA for the Hall bar).

The dependence of DC RDNMR signals of both samples on different cool-downs is shown in Supplementary Fig. 2, where the signal amplitude of the Hall bar is found to depend on different cool-downs while that of the Corbino disk does not. We assign the edge transport in the Hall bar to account for this difference: a random impurity potential in different cool-downs will modify the local chemical levels near DW that leads to a change in the transmission probability for the electrons to pass along or across the DW[8] and thus in $\Delta\sigma_{xx}/\sigma_{xx}^{\text{sat}}$. Furthermore, it is found that the direction of current flow and the orientation of magnetic field change the signal amplitude of the Hall bar but do not affect that of the Corbino disk (Fig. 2b and Supplementary Fig. 1). This difference is also induced by the edge transport. It is shown in Fig. 3a,b,g,h that the direction of current flow and the orientation of magnetic field determine the preferred path of edge current. Different paths may lead to a difference in the transmission probability for the electrons to pass across the DW in bulk, which accounts for the difference in the amplitude of bulk signals.

The RDNMR signal obtained from the AC measurement is quite different from the DC counterpart for both Corbino disk and Hall bar. It is seen from Supplementary Fig. 3 that there is no AC signal in the Corbino disk, while a weak AC signal is present at $B$ = 12 T but absent at $B$ = -12 T in the Hall bar. These results provide support for the edge mode of the Hall bar as discussed below. The AC current with opposite flow directions is believed to suppress the bulk DNP of the Hall bar (see main text). However, the direction of edge current flow depending on the orientation of $B$ cannot be changed by the AC current, making the polarized nuclei with opposite spins distribute along the length rather than the width of DW. The spatial overlap between the two species varies with the DW length that may be modified by external factors (magnetic field, cool-down, etc.), which accounts for diverse RDNMR responses (one example is given in Supplementary Fig. 3).

**Supplementary References**

1. McEuen P. L., Szafer A., Richter C. A., Alphenaar B. W., Jain J. K., Stone A. D., Wheeler R. G. & Sacks R. N. New resistivity for high-mobility quantum Hall conductors. *Phys. Rev. Lett.* **64**, 2062-2065 (1990).

2. Ramvall P., Carlsson N., Omling P., Samuelson L., Seifert W., Wang Q., Ishibashi K. & Aoyagi Y. Quantum transport in high mobility modulation doped $Ga_{0.25}In_{0.75}As$/InP quantum wells. *J. Appl. Phys.* **84**, 2112-2122 (1998).

3. Liu H. W., Yang K. F., Mishima T. D., Santos M. B. & Hirayama Y. Dynamic nuclear polarization and nuclear magnetic resonance in the simplest pseudospin quantum Hall ferromagnet. *Phys. Rev. B* **82**, 241304(R) (2010).

4. Brey L. & Tejedor C. Spins, charges, and currents at domain walls in a quantum Hall Ising ferromagnet. *Phys. Rev. B* **66**, 041308(R) (2002).

5. Jungwirth T. & MacDonald A. H. Resistance spikes and domain wall loops in Ising quantum Hall ferromagnets. *Phys. Rev. Lett.* **87**, 216801 (2001).

6. Shklovskii B. & Efros A. *Electronic properties of doped semiconductors* (Springer-Verlag, Berlin, 1984).

7. Fal'ko V. I. & Iordanskii S. V. Topological defects and Goldstone excitations in domain walls between ferromagnetic quantum Hall liquids. *Phys. Rev. Lett.* **82**, 402-405 (1999).

8. Mitra, A. & Girvin, S. M. Electron/nuclear spin domain walls in quantum Hall systems. *Phys. Rev. B* **67**, 245311 (2003).